\documentclass[conference]{IEEEtran}
\usepackage{cite}
\usepackage{amsmath,amssymb,amsfonts}
\usepackage{graphicx}
\usepackage{xcolor}
\usepackage{subfig}
\usepackage{multirow}
\usepackage{booktabs}
\usepackage{tcolorbox}
\usepackage{stfloats}
\usepackage[ruled,linesnumbered]{algorithm2e}

\renewcommand{\sf}[1]{\textsf{#1}}
\newcommand{\nmodels}{\not\mathrel|\joinrel=}

\newcommand{\eventually}[0]{\Diamond}
\newcommand{\globally}[0]{\Box}
\newcommand{\estrobustness}[0]{\widehat{\mathbf{\Delta}}}

\newtheorem{definition}{Definition}[section]
\newtheorem{problem}{Problem}[section]

\begin{document}

\title{Investigating Robustness in Cyber-Physical Systems: Specification-Centric Analysis in the face of System Deviations}

\author{
\IEEEauthorblockN{Changjian Zhang\IEEEauthorrefmark{1}}
\IEEEauthorblockA{\textit{Carnegie Mellon University} \\
Pittsburgh, PA USA \\
changjiz@andrew.cmu.edu}
\and
\IEEEauthorblockN{Parv Kapoor\IEEEauthorrefmark{1}}
\IEEEauthorblockA{\textit{Carnegie Mellon University} \\
Pittsburgh, PA USA \\
parvk@andrew.cmu.edu}
\and
\IEEEauthorblockN{R\^omulo Meira-G\'oes}
\IEEEauthorblockA{\textit{The Pennsylvania State University} \\
State College, PA USA \\
romulo@psu.edu}
\and
\IEEEauthorblockN{David Garlan}
\IEEEauthorblockA{\textit{Carnegie Mellon University} \\
Pittsburgh, PA USA \\
dg4d@andrew.cmu.edu}
\and
\IEEEauthorblockN{Eunsuk Kang}
\IEEEauthorblockA{\textit{Carnegie Mellon University} \\
Pittsburgh, PA USA \\
eunsukk@andrew.cmu.edu}
\and
\IEEEauthorblockN{Akila Ganlath}
\IEEEauthorblockA{\textit{Toyota InfoTech Labs} \\
Mountain View, CA USA \\
akila.ganlath@toyota.com}
\and
\IEEEauthorblockN{Shatadal Mishra}
\IEEEauthorblockA{\textit{Toyota InfoTech Labs} \\
Mountain View, CA USA \\
shatadal.mishra@toyota.com}
\and
\IEEEauthorblockN{Nejib Ammar}
\IEEEauthorblockA{\textit{Toyota InfoTech Labs} \\
Mountain View, CA USA \\
nejib.ammar@toyota.com}
}

\maketitle

\begingroup
\renewcommand\thefootnote{\IEEEauthorrefmark{1}}
\footnotetext{Both authors contributed equally to this research.}
\endgroup

\thispagestyle{plain}
\pagestyle{plain}

\begin{abstract}
The adoption of cyber-physical systems (CPS) is on the rise in complex physical environments, encompassing domains such as autonomous vehicles, the Internet of Things (IoT), and smart cities. A critical attribute of CPS is robustness, denoting its capacity to operate safely despite potential disruptions and uncertainties in the operating environment. This paper proposes a novel specification-based robustness, which characterizes the effectiveness of a controller in meeting a specified system requirement, articulated through Signal Temporal Logic (STL) while accounting for possible deviations in the system. This paper also proposes the robustness falsification problem based on the definition, which involves identifying minor deviations capable of violating  the specified requirement. We present an innovative two-layer simulation-based analysis framework designed to identify subtle robustness violations. To assess our methodology, we devise a series of benchmark problems wherein system parameters can be adjusted to emulate various forms of uncertainties and disturbances. Initial evaluations indicate that our falsification approach proficiently identifies robustness violations, providing valuable insights for comparing robustness between conventional and reinforcement learning (RL)-based controllers.
\end{abstract}

\section{Introduction}
\emph{Robustness} is a type of  property that characterizes the ability of a system to function correctly in the presence of uncertainties in the environment. Robustness plays an especially important role in modern cyber-physical systems (CPS) that operate in dynamic and uncertain environments, such as autonomous vehicles, medical devices, the Internet of Things (IoT), and smart cities. The mission-critical and safety-critical nature of CPS accentuate the need to provide a high level of robustness, as a failure to do so could result in severe consequences, from safety hazards to economic losses.

The concept of robustness has received extensive attention within the control theory literature~\cite{robust-control-parametric,Zhou1998-essential-robust-control}. More recently, with the increasing popularity of reinforcement learning (RL)-based controllers, robust RL \cite{Moos-RobustRL,xu2022trustworthy} has also become an active research topic. These lines of research investigate how to design and analyze a controller that is \emph{robust}, in that it is capable of maintaining desired system behavior in the presence of possible system \emph{deviations}---environmental uncertainties, observation or actuation errors, disturbances, and modeling errors. In traditional control, the notion of ``desired behavior'' is typically captured by stability properties \cite{Zames1996-IO-robustness,Sontag2008-stability,Tabuada2016-robustness}, stating that the controller is able to maintain a system output within a certain bound (given bounded changes in input). In RL, the desired behavior is expressed using a reward function \cite{Moos-RobustRL,xu2022trustworthy}; i.e., the system achieves a desired level of reward despite external disturbances. 

One limitation of the existing definitions of robustness is the limited expressiveness over the notion of desired system behavior to be maintained in the presence of deviations. While certain system requirements (e.g., invariance) can be encoded as a constraint over a system output, other types of requirements, especially those that capture time-varying behavior (e.g., ``the vehicle must come to a stop in the next 3 seconds"), cannot be directly encoded in this manner. Similarly, it is well-known that encoding a high-level system requirement using a reward function is a challenging task that requires a significant amount of domain expertise and manual effort via reward shaping \cite{10.5555/645528.657613, Booth_Knox_Shah_Niekum_Stone_Allievi_2023}.

%Most of these works focus on building a robust controller with respect to a certain type of uncertainty, whereas a concrete definition of robustness that captures the ability of a controller to ensure the system satisfies a system specification under uncertainties is missing\ek{I still don't find this statement regarding ``a concrete definition" not very satisfying or compelling. Let's talk more about this tomorrow}\rmg{I don't quite agree with this statement since the definition of robust control margin is a concrete definition of robustness.}\cj{If we emphasize the use of system specification in STL, would that make a difference?}. Therefore, different from building robust controllers, we instead focus on the evaluation perspective of robustness in this work. 

In this paper, we propose \emph{specification-based robustness}, a new notion of robustness that allows the engineer to directly specify a high-level system requirement as desired behavior to be maintained. Our definition assumes a \emph{parametric} representation of a system, where \emph{system parameters} capture the dynamics of the system (e.g., acceleration of a nearby vehicle) that may be affected by possible deviations (e.g., sensor errors). A system is initially assigned a set of nominal parameters that describe its expected dynamics. Then, a change in one or more parameters, denoted by $\delta$, corresponding to a possible deviation that may occur, is put into effect. 
 Finally, a controller is said to be \emph{robust} with respect to a given specification---specified using \emph{Signal Temporal Logic (STL)}~\cite{stl}---and a set of deviations if and only if the controller is capable of \emph{satisfying} the specification even under those deviations. STL is particularly well-suited as the notation for specifying CPS requirements, as it can be used to capture a wide range of temporal behavior over continuous state variables. 
%We propose a formal definition of behavioral robustness\ek{should define what "behavioral" means in this context} for CPS controllers where the uncertainties are captured by deviations in system parameters and the system specifications are modeled in signal temporal logic (STL)\cite{stl}. 
%Specifically, \emph{system parameters} are parameters that affect the system dynamics (e.g., mass of a vehicle). The deviations in these parameters can represent a variety of uncertainties or errors (e.g., environmental uncertainties, observation or actuation errors) that would lead to system dynamic changes. 
%Then, we consider only safety specifications in STL. Finally, robustness is defined as the set of all deviations in system parameters such that the system under control still satisfies a desired STL safety property.

Based on this robustness definition, we propose a new type of analysis problem called the \emph{robustness falsification}. The goal of robustness falsification is to find deviations in system parameters that may result in a violation of the desired system specification. Specifically, we argue that identifying a violation closer to the nominal system parameters would be more valuable, since such a violation is more likely to occur in practice. Intuitively, our system needs to be robust to these deviations before addressing the ones that are further away from the nominal set.  These identified violations could then be used to redesign or re-train the controller for improved robustness, or to build a run-time monitor to detect when the systems' parameters deviate into an unsafe region. 

In addition, we propose a novel simulation-based framework where the falsification problem is formulated as a two-layer optimization problem. In the basic layer, for a given system deviation $\delta$ (representing a particular system dynamics), an optimization-based method is used to find a falsifying signal; i.e., a sequence of system states that results in a violation of the given STL specification. In the upper layer, the space of possible deviations is explored to find small deviations that result in a specification violation, repeatedly invoking the lower-layer falsification method. The information generated from the lower layer (i.e., the degree of specification violation by a signal) is, in turn, used to guide the search towards smaller deviations that are also likely to result in a violation.

To evaluate the effectiveness of our falsification approach, we have constructed a set of benchmark caes studies that contain both traditional and RL-based controllers. In particular, these benchmark systems are configurable with system parameters to generate a range of systems with varying dynamics, to enable the type of robustness analysis that we propose in this paper. Our preliminary evaluation shows that our approach can be used to effectively find small deviations that cause a specification violation in these systems. Furthermore, the results from our analysis offer insights about the comparative robustness of controllers that are based on classical control methods versus RL. 

%Furthermore, we have found that CPS case studies used in academia often do not provide built-in support for configurable system parameters for changing system dynamics. Therefore, we present a CPS robustness benchmark where we provide several systems with various types of configurable parameters, adopted from existing Python and Matlab case studies. We also provide various classic and RL controllers for these systems to evaluate their robustness. Finally, we conduct an evaluation of our falsification approach over this benchmark.

This paper makes the following contributions:
\begin{itemize}
    \item We present a novel, formal definition of \emph{specification-based robustness} for CPS (Section~\ref{sec:robustness}).
    \item We propose the \emph{robustness falsification problem} (Section~\ref{sec:falsification}) and a two-layer optimization-based method for finding small violating deviations (Section~\ref{sec:framework}).
    \item We present a CPS robustness benchmark (Section~\ref{sec:benchmark}) and the results from experimental evaluation of our falsification approach over the benchmark
    (Section~\ref{sec:evaluation}).
\end{itemize}

% The rest of this paper is organized as follows: Section \ref{sec:background} provides the necessary formal background for discrete-time control systems and STL. Section \ref{sec:robustness} presents our formal definition of robustness, and Section \ref{sec:falsification} presents the definition of the robustness falsification problem. Then, Section \ref{sec:framework_benchmark} presents our analysis framework and the CPS robustness benchmark, and Section \ref{sec:evaluation} presents our evaluation results. Finally, Section \ref{sec:related_work} summarizes the related work on CPS robustness and Section \ref{sec:conclusion} discusses the limitation and the future work and concludes the paper.

\section{Preliminaries}\label{sec:background}
In this work, we consider CPS in discrete-time stochastic systems and use \emph{Signal Temporal Logic (STL)}~\cite{stl}  to specify system properties. This section presents the necessary background on these concepts.

\subsection{Discrete-time Stochastic Systems}
A \emph{discrete-time stochastic system} $\mathbf{S}$ is defined by
\begin{align*}
    x_0 \sim h(\cdot), \quad x_{t+1} \sim f(x_t, u_t)
\end{align*}
where $t \in \mathbb{N}$ is the time indices, $x_t, x_{t+1} \in \mathbf{X} \subseteq \mathbb{R}^n$ are the states of the system at time $t$ and $t+1$, respectively, $u_t \in \mathbf{U} \subseteq \mathbb{R}^m$ is the control input at time $t$, $x_0$ is the initial state from the distribution $h$, and $f$ defines the probability distribution of the system transiting to state $x_{t+1}$ given state $x_t$ and control input $u_t$, i.e., $P(x_{t+1}~|~x_t, u_t)$.

Then, we consider a \emph{deterministic controller} $\mathbf{C}$ of the form 
$$g: Seq(\mathbf{X}) \to \mathbf{U}$$
where $Seq(\mathbf{X})$ is a sequence of system states. Then, given an initial state $x_0$, a trajectory of the system is defined as
\begin{align*}
    \sigma = (x_0 \xrightarrow{u_0} x_1 \ldots x_i \xrightarrow{u_i} x_{i+1} \ldots)
\end{align*}
where $u_i = g(x_0, \ldots, x_i)$ and $x_{i+1} \sim f(x_i, u_i)$.
% A \emph{memoryless} controller is defined as $g': \mathbf{X} \to \mathbf{U}$ such that the control action only relates to the current state of the system.
Finally, we use $\mathcal{L}(\mathbf{S}||\mathbf{C})$ to represent the behavior of the controlled system, i.e., it is the set of all trajectories of a system $\mathbf{S}$ under the control of $\mathbf{C}$.
% In addition, we assume $g$ has full observability over $\mathbf{X}$ in this paper.

\subsection{Signal Temporal Logic}\label{sec:stl}
A signal $\mathbf{s}$ is a function $\mathbf{s}: T \to D$ that maps a time domain $T \subseteq \mathbb{R}_{\geq 0}$ to a $k$ real-value space $D \subseteq \mathbb{R}^k$, where $\mathbf{s}(t) = (v_1, \ldots, v_k)$ represents the value of the signal at time $t$. Then, an STL formula is defined as:
$$\phi := \mu ~|~ \neg \phi ~|~ \phi \land \psi ~|~ \phi \lor \psi ~|~ \phi ~\mathcal{U}_{[a,b]}~\psi$$
where $\mu$ is a predicate of the signal $\mathbf{s}$ at time $t$ in the form of $\mu \equiv \mu(\mathbf{s}(t)) > 0$ and $[a, b]$ is the time interval (or simply $I$). The \emph{until} operator $\mathcal{U}$ defines that $\phi$ must be true until $\psi$ becomes true within a time interval $[a, b]$. Two other operators can be derived from \emph{until}: \emph{eventually} ($\eventually_{[a,b]}~\phi := \top~\mathcal{U}_{[a,b]}~\phi$) and \emph{always} ($\globally_{[a,b]}~\phi := \neg\eventually_{[a,b]}~\neg\phi$).

% \subsection{Satisfaction Metric of STL}
The satisfaction of an STL formula can be measured in a quantitative way as a real-valued function $\rho(\phi, \mathbf{s}, t)$, which represents the difference between the actual signal value and the expected one \cite{stl}. For example, given a formula $\phi \equiv \mathbf{s}(t) - 3 > 0$, if $\mathbf{s} = 5$ at time $t$, then the satisfaction of $\phi$ can be evaluated by $\rho(\phi, \mathbf{s}, t) = \mathbf{s}(t) - 3 = 2$. The complete definition of $\rho$ is defined recursively as follows:
\begin{align*}
    \rho(\mu, \mathbf{s}, t) & = \mu(\mathbf{s}(t)) \\
    \rho(\neg \phi, \mathbf{s}, t) & = - \rho(\phi, \mathbf{s}, t) \\
    \rho(\phi \land \psi, \mathbf{s}, t) & = \min \{ \rho(\phi, \mathbf{s}, t), \rho(\psi, \mathbf{s}, t) \} \\
    \rho(\phi \lor \psi, \mathbf{s}, t) & = \max \{ \rho(\phi, \mathbf{s}, t), \rho(\psi, \mathbf{s}, t) \} \\
    \rho(\globally_{I}~\phi, \mathbf{s}, t) & = \inf_{t' \in I + t} \rho(\phi, \mathbf{s}, t') \\
    \rho(\eventually_{I}~\phi, \mathbf{s}, t) & = \sup_{t' \in I + t} \rho(\phi, \mathbf{s}, t') \\
    \rho(\phi~\mathcal{U}_{I}~\psi, \mathbf{s}, t) & = \sup_{t_1 \in I + t} \min \{ \rho(\psi, \mathbf{s}, t_1), \inf_{t_2 \in [t, t_1]} \rho(\phi, \mathbf{s}, t_2) \}
\end{align*}

In literature, the value computed by the function $\rho(\cdot)$ is often referred as the \emph{robustness} value of an STL formula $\phi$. However, to distinguish it from our proposed robustness definition, we will use the term \emph{STL satisfaction value}  for the rest of the paper.

\section{Motivating Example}\label{sec:motivation}
\begin{figure}[!ht]
\centering
\includegraphics[width=\linewidth]{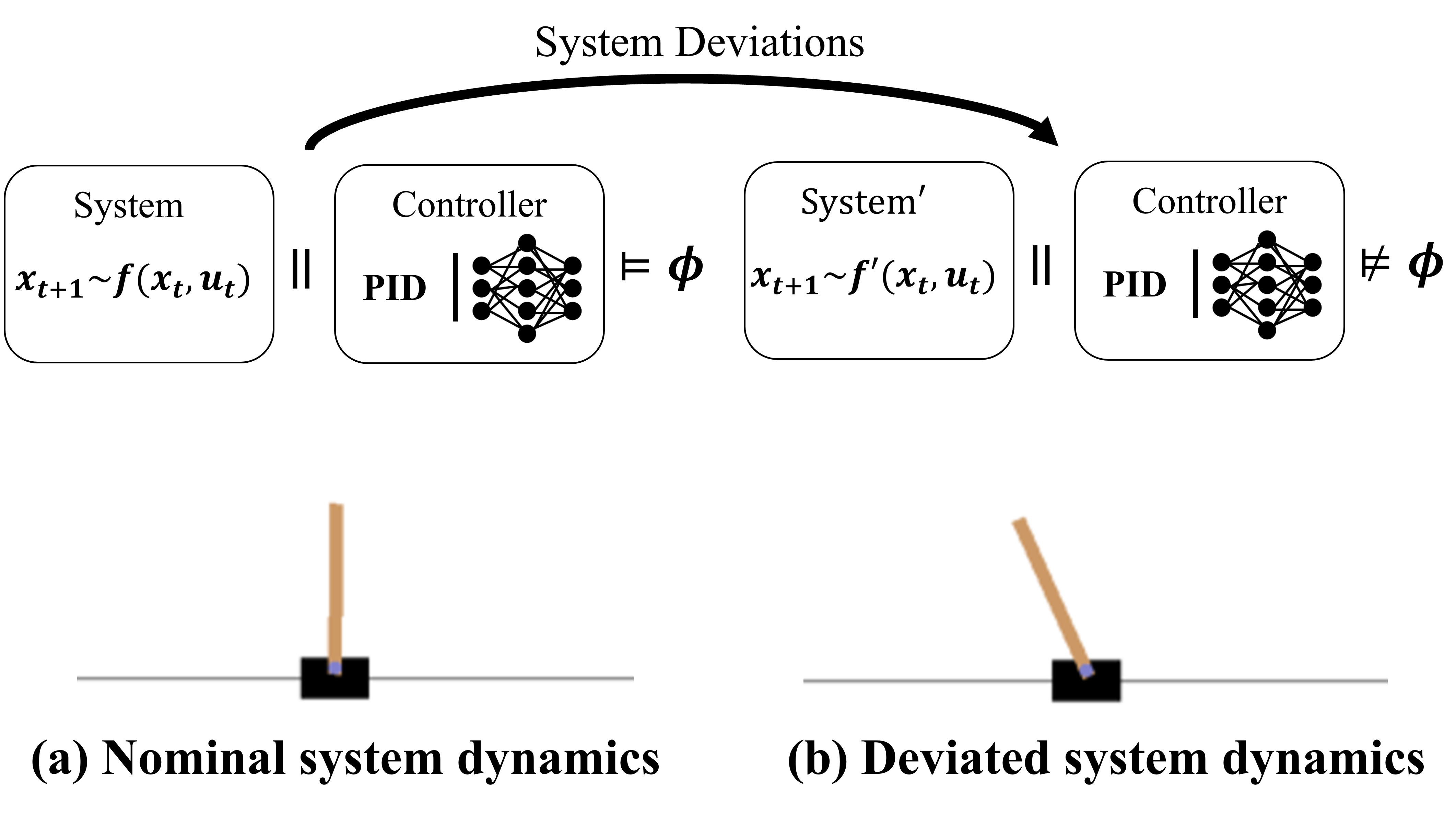}
\caption{Behavior of the Cart-Pole system under different system parameters, where the two systems have different transition function $f$ and $f'$, respectively.}
\label{fig:cart-pole-example}
\end{figure}

In this section, we use the \emph{Cart-Pole} system\footnote{https://www.gymlibrary.dev/environments/classic\_control/cart\_pole/}, depicted in Figure~\ref{fig:cart-pole-example}, to illustrate our robustness definition and analysis. 
The Cart-Pole system describes a pendulum attached to a cart moving along a frictionless track.
The control objective for this system is to maintain the pendulum upright by pushing the cart left or right with a fixed force \cite{cart-pole}. 
This objective is defined by a safety specification: \emph{the angle of the pendulum $\theta$ and the distance of the cart from the origin $x$ should both be within certain thresholds.} Formally, it can be specified by an STL formula: $$\Box (|\theta| < C_1 \land |x| < C_2)$$ where $C_1$ and $C_2$ are the constant thresholds for the angle and the distance, respectively.

Given the Cart-Pole dynamics, we can design a PID controller or train an RL agent such that the controlled system satisfies the safety specification above. 
However, the actual system dynamics may deviate from the \emph{nominal} Cart-Pole dynamics due to modeling errors (e.g., a change in the mass of the cart) or actuation errors (e.g., a change in the force applied to the cart). 
% However, even though the controller can guarantee the safety property given the expected system conditions, the actual system dynamics may deviate due to modeling deviations (e.g., the mass of the cart changes) or actuation errors (e.g., the force applied to the cart changes). 
Such deviations in dynamics may cause this controlled system to violate the safety property.
% In this case, we say the controller is not \emph{robust} against such deviations.

% \begin{figure}[!ht]
% \centering
% \subfloat[Nominal system dynamics]{
%     \includegraphics[width=0.4\linewidth]{figures/Rob_figure_3.pdf}
% }
% \hfill
% \subfloat[Deviated system dynamics]{
%     \includegraphics[width=0.4\linewidth]{figures/Rob_figure_2.pdf}
% }
% \caption{Behavior of the Cart-Pole system under different system dynamics.}
% \label{fig:cart-pole-example}
% \end{figure}

To capture such deviations in system dynamics, we model the system as being \emph{parametric}  with two \emph{system parameters}, $m$ (mass, in $kg$) and $F$ (force, in $N$). 
Figure~\ref{fig:cart-pole-example} shows the behavior of the Cart-Pole system under different dynamic conditions. 
In Figure~\ref{fig:cart-pole-example}(a), the controller is deployed in the nominal Cart-Pole dynamics ($m = 1, F = 10$).
In this scenario, the pendulum is always maintained in the upright position, i.e., the safety specification is satisfied.
In Figure~\ref{fig:cart-pole-example}(b), we show the same controller deployed under a deviated Cart-Pole dynamics ($m=1, F=20$). 
In this scenario, a larger force is applied to the cart while pushing it from the left.
This deviation makes the pendulum swing at a larger angle than expected, violating the safety property, and eventually falling onto the ground.

The engineer of such a system may wish to ask questions like: \emph{How robust is my controller against possible deviations in the system?} More specifically, \emph{how much change in the system dynamics can the controller tolerate before it begins to violate the given safety specification?} We formulate this question as a type of analysis problem called \emph{robustification falsification}, where the goal is to find deviations in system parameters (e.g., the changes in the mass and force of the Cart-Pole) where the system with the resulting dynamics may violate the given specification. In general, it is relatively straightforward to find such a violation by picking a large deviation value (e.g., minimal or maximum possible value for the mass and force), but deviations that are closer to the nominal parameters (e.g., original mass and force) are more likely to occur in practice, and  are also more valuable for the engineer. Thus, our falsification process attempts to find violations with \emph{small} deviations; i.e., minimal parameter changes that introduce a risk of specification violation into the system.
%In general, there are an infinite number of possible deviations in system parameters, and computing a complete range of deviations against which the system is robust would be computationally intractable. Therefore, in this work, we propose to solve the problem of finding deviations that would lead to a safety violation; in other words, a falsification problem. 
The output of this analysis (i.e., violations) can help the engineer identify brittleness in the controller, and can be used to redesign or retrain the controller to improve its robustness. In the next section, we present the formal definition of  specification-based robustness and the robustness falsification problem.

\section{Robustness Definition}\label{sec:robustness}

\subsection{Definition of Specification-Based Robustness}
As shown in the motivating example, in this work, we use STL to specify the desired properties of a system, and system parameters to capture the deviations in system dynamics. Parameters can represent a variety of deviations such as environmental disturbances (e.g., wind or turbulence),  internal deviations (e.g., mass variation of a vehicle), observation errors (e.g., sensor errors), or actuation errors (e.g., errors in steering control). Then, to capture systems with such varying dynamics by using parameters, we leverage the notion of \emph{parametric control systems} \cite{Bhattacharyya:1995,weinmann:2012uncertain}.

A \emph{parametric} discrete-time stochastic system $\mathbf{S}^\Delta$ defines a set of systems such that $\Delta \subseteq \mathbb{R}^k$ represents the parameter domain, and for any $\delta \in \Delta$, an instance of a parametric system $\mathbf{S}^\delta$ has the form
\begin{align*}
    x_0 \sim h^{\delta}(\cdot), \quad x_{t+1} \sim f^{\delta}(x_t, u_t)
\end{align*}
where the initial state distribution $h^{\delta}$ and the state transition distributions $f^{\delta}$ are both defined by the parameter $\delta$.

% \ek{We need to say more about what assumptions we make about $\delta$. Is it just limited to multi-dimensional vectors of real numbers, or can be it something more general? In general, since we want to be able to compare different $\delta$'s and also compute distance between them, I assume $\delta$ comes from a metric space.} \cj{It could be but our evaluation cannot show that. Since in our case studies, we only have deviations in real numbers applied to the environment.} \ag{I don't have issue with it, but I am curious why we are saying the initial state distribution and dynamics are both parameterized by $\delta$. Is this for convenience or is there some deeper reason?} \cj{One reason is to keep it general enough, because we find there are some environments like in OpenAI Gym the initial state in computed by, e.g., applying a force to the robot. So if we change the system parameters, the initial state distribution will also change because it is computed not predefined.}

Parameter $\delta$ represents a deviation to a system and $\Delta$ represents the domain of all deviations of interest. In addition, we use $\delta_0 \in \Delta$ to represent the zero-deviation point, i.e., the parameter under which the system $\mathbf{S}^{\delta_0}$ exhibits the expected, normative behavior. Then, we define a system as being robust against a certain deviation as follows:
\begin{definition}\label{def:robust-system}
For a system $\mathbf{S}$, a controller $\mathbf{C}$, a deviation parameter $\delta$, and an STL property $\phi$, we say the system is \emph{robust} against the deviation when the parametric form of  $\mathbf{S}$ with parameter $\delta$ under the control of  $\mathbf{C}$ satisfies the property, i.e., $\mathbf{S}^\delta || \mathbf{C} \models \phi$.
\end{definition}

Then, the robustness of a controller can be defined as all the possible deviations which the system is robust against. Formally, we have,
\begin{definition}\label{def:robustness}
For a system $\mathbf{S}$, a controller $\mathbf{C}$, and an STL property $\phi$, the robustness of the controller is defined as the \textbf{maximal} $\mathbf{\Delta} \subseteq \mathbb{R}^k$ s.t. $\forall \delta \in \mathbf{\Delta} : \mathbf{S}^{\delta}||\mathbf{C} \models \phi$.
\end{definition}
% \rmg{Do we need to be more formal here of what we mean by maximal? How are we comparing the real vectors? Once we have that, should we use sup instead of maximal? Unless we always bounding $R^k$}\cj{Good question, actually I don't know how to compare two sets of real vectors. Can we keep it informal, because we don't compute it anyway?} \rmg{It depends, people use norms, like infinity norm, or direct comparison. For example, $[a;b]\geq [c;d]$ if $a\geq c$ and $b\geq d$. I believe we are using more the second case. }\cj{But in an extreme case, robustness could include two separate sets. I feel we can stay abstract here because a concrete definition would assume a shape of robustness}

In other words, the robustness of a controller $\mathbf{C}$ is measured by the maximal parameter range $\mathbf{\Delta}$ of a system where each deviated system $\mathbf{S}^\delta$ of it still satisfies the property under the control of $\mathbf{C}$.

\subsection{Strict Satisfaction of Robustness}
Definition \ref{def:robust-system} and \ref{def:robustness} depend on the interpretation of $\mathbf{S}^{\delta} || \mathbf{C} \models \phi$, i.e., the system is robust when the STL property is satisfied. However, STL is evaluated against a certain trajectory. From the literature \cite{Corso2021-cps-survey}, the most common interpretation is that \emph{a system must not contain a trajectory that violates the STL property}. In other words, even in the worst scenario that is less likely to occur in a stochastic system, it should still guarantee the property. This interpretation enforces a strong guarantee of the system, and thus we call it the \emph{strict} satisfaction of STL in this work. Formally, we have,
\begin{definition}\label{def:strict-stl-satisfaction}
A discrete-time stochastic system $\mathbf{S}$ \emph{strictly} satisfies an STL property $\phi$ under the control of a controller $\mathbf{C}$ iff every controlled trajectory produces a non-negative STL satisfaction value, i.e.,
\begin{align*}
    \mathbf{S}||\mathbf{C} \models \phi \Leftrightarrow \forall \sigma \in \mathcal{L}(\mathbf{S}||\mathbf{C}) : \rho(\phi, \mathbf{s}_{\sigma}, 0) \geq 0
\end{align*}
where $\mathbf{s}_{\sigma}$ is the signal of state values of trajectory $\sigma$.
\end{definition}

With this interpretation, we can then restate Definition \ref{def:robustness} as:
\begin{definition}\label{def:strict-robustness}
The robustness of a controller that \emph{strictly} satisfies an STL property $\phi$ is the maximal $\mathbf{\Delta}$ s.t. $$\forall \delta \in \mathbf{\Delta}, \sigma \in \mathcal{L}(\mathbf{S}^{\delta}||\mathbf{C}) : \rho(\phi, \mathbf{s}_{\sigma}, 0) \geq 0$$
\end{definition}
This  definition delineates a strong robustness guarantee of a controller with respect to a given system specification.

\section{Robustness Analysis}\label{sec:falsification}
\subsection{Robustness Falsification}
According to Definition \ref{def:strict-robustness}, to compute the robustness of a controller, we need to: (1) (formally) show that a stochastic system does not contain a trajectory that violates the STL property, and (2) computing the maximal parameter set $\Delta$, which could be in any (non-convex) or even non-continuous shape, where all system instances $S^\delta$ should satisfy step (1). Apparently, such process would be computationally intractable.

Therefore, in this work, instead of computing or approximating the robustness $\mathbf{\Delta}$, we consider the problems of falsifying a given estimation of robustness $\estrobustness$, i.e., finding a deviation $\delta \in \estrobustness$ such that the system is not robust given the control agent. More formally, we define
\begin{problem}[Robustness Falsification]\label{prob:robustness-falsify}
For a system $\mathbf{S}$, a controller $\mathbf{C}$, and an STL property $\phi$, given a robustness estimation $\estrobustness \subseteq \mathbb{R}^k$, the goal of a \emph{robustness falsification problem} $\mathcal{F}(\mathbf{S}, \mathbf{C}, \phi, \estrobustness)$ is to find a deviation $\delta \in \estrobustness$ s.t. $\exists \sigma \in \mathcal{L}(\mathbf{S}^\delta || \mathbf{C}): \rho(\phi, \mathbf{s}_{\sigma}, 0) < 0$.
\end{problem}

% \begin{problem}[Robustness Falsification]\label{prob:robustness-falsify}
% For a system $\mathbf{S}$, a controller $\mathbf{C}$, and an STL property $\phi$, given a robustness estimation $\estrobustness \subseteq \mathbb{R}^k$, the goal of a \emph{robustness falsification problem} $\mathcal{F}(\mathbf{S}, \mathbf{C}, \phi, \estrobustness)$ is to find a deviation $\delta \in \estrobustness$ s.t. $\mathbf{S}^\delta || \mathbf{C} \nmodels \phi$.
% \end{problem}

% Then, if we employ the worst-case robustness definition (Definition \ref{def:worst-case-satisfaction}), the problem can be restated as:
% \begin{problem}[Worst-Case Robustness Falsification]
% The goal of a \emph{robustness falsification problem} $\mathcal{F}(\mathbf{S}, \mathbf{C}, \phi, \estrobustness)$ is to find a deviation $\delta \in \estrobustness$ s.t. $\exists \sigma \in \mathcal{L}(\mathbf{S}^\delta || \mathbf{C}): \rho(\phi, \mathbf{s}_{\sigma}, 0) < 0$.
% \end{problem}\ek{I am not entirely sure this Problem 4.2 is fundamentally different from Problem 4.1. Is 4.2 simply restating 4.1?}\cj{It is restating using the worst-case STL satisfaction.}
% \rmg{This probably is related to my confusion of having an abstract $\models$ and then a ``worst-case satisfaction".}

\subsection{Minimum Robustness Falsification}
Intuitively, a larger deviation (i.e., a deviation parameter that is far away from the expected, normative system parameter) would likely cause a larger deviation in the system behavior leading to a specification violation. Thus, an arbitrary large deviation leading to a violation may not be useful. Instead, we consider the \emph{minimum} deviation.
\begin{problem}\label{prob:min-robustness-falsify}
Given a \emph{minimum} robustness falsification problem $\mathcal{F}_{min}(\mathbf{S}, \mathbf{C}, \phi, \estrobustness)$, let $\delta_0 \in \estrobustness$ be the zero-deviation point, the goal is to find a deviation $\delta \in \estrobustness$ s.t. $\mathbf{S}^\delta || \mathbf{C} \nmodels \phi$ and $\delta$ minimizes a distance measure $\lVert \delta - \delta_0 \rVert_p$.
\end{problem}

We argue that this is a more reasonable evaluation in practice because it is crucial to ensure that the system is robust in the expected environment and the environment that has small deviations that are more likely to occur in real operation conditions compared to arbitrary large deviations.

\subsection{Falsification by Optimization}\label{sec:falsification-by-optimization}
Since an STL formula can be quantitatively measured, the robustness falsification problems can be them formulated as optimization problems. Consider a real-valued system evaluation function $\Gamma(\mathbf{S}, \mathbf{C}, \phi) \in \mathbb{R}$, where when its value is smaller than zero, the controlled system violates the property, i.e., $$\Gamma(\mathbf{S}, \mathbf{C}, \phi) < 0 \Leftrightarrow \mathbf{S}||\mathbf{C} \nmodels \phi$$
and the smaller the value, the larger the degree the property is violated.

Then, robustness falsification problem $\mathcal{F}(\mathbf{S}, \mathbf{C}, \phi, \estrobustness)$ can be formulated as the following optimization problem:
\begin{align}\label{eq:any_deviation}
    \mathop{\arg\min}\limits_{\delta \in \estrobustness}~\Gamma(\mathbf{S}^{\delta}, \mathbf{C}, \phi)
\end{align}
i.e., by finding a parameter $\delta \in \estrobustness$ that minimizes the evaluation function $\Gamma$, if the function value is negative, then this parameter $\delta$ indicates a deviation where the system violates the property $\phi$.

Specifically, in the case of strict satisfaction of robustness, the system evaluation function $\Gamma$ is defined as
\begin{align}\label{eq:system_evaluation}
    \Gamma(\mathbf{S}, \mathbf{C}, \phi) = \min \{ \rho(\phi, \mathbf{s}_{\sigma}, 0)~|~\sigma \in \mathcal{L}(\mathbf{S}||\mathbf{C}) \}
\end{align}
Note that, the problem of finding the minimum STL satisfaction value of a system has also been applied in CPS falsification in the literature \cite{Corso2021-cps-survey}.

Finally, we can formulate a minimum robustness falsification problem $\mathcal{F}_{min}(\mathbf{S}, \mathbf{C}, \phi, \estrobustness)$ as a constrained optimization problem:
\begin{align}\label{eq:min_deviation}
    \mathop{\arg\min}\limits_{\delta \in \estrobustness}~\lVert \delta - \delta_0 \rVert_p ~ s.t. ~ \Gamma(\mathbf{S}^{\delta}, \mathbf{C}, \phi) < 0
\end{align}

\section{Simulation-Based Robustness Analysis Framework}\label{sec:framework}
As CPS grow in their complexity and scale, it becomes hard to construct white-box mathematical models of their dynamics. Additionally, RL controllers are gaining widespread popularity and applicability in the CPS domain; however, it is known to be hard to formally model RL controllers, especially those using deep neural networks (DNNs). Therefore, in this work, we implement a simulation-based analysis framework to solve the robustness falsification problems, i.e., Problem \ref{prob:robustness-falsify} and \ref{prob:min-robustness-falsify}. The framework supports analysis for black-box CPS and controllers (either classical controllers or RL controllers) and solving the robustness falsification problems using off-the-shelf sampling and optimization algorithms and tools.

% \begin{figure}
%     \centering
%     \includegraphics[width=\linewidth]{figures/fig-algorithm.png}
%     \caption{Overview of the robustness falsification process.}
%     \label{fig:algorithm}
% \end{figure}

\begin{algorithm}
\SetKwFunction{Instantiate}{Instantiate}
\SetKwFunction{CPSFalsification}{CPSFalsification}
\SetKwFunction{NextCandidates}{NextCandidates}
\SetKwFunction{UpdateBest}{UpdateBest}
\SetKwInOut{Input}{Input}
\SetKwInOut{Output}{Output}
\Input{$\mathbf{S}, \mathbf{C}, \phi, \estrobustness$, and objective function $f$}
\Output{violation $\delta_{best} \in \estrobustness$}

$\delta_{best} \gets nil$\;
$X \gets~\text{random values in}~\estrobustness$ \;
\While{termination criteria = false}{
    $V \gets \langle \rangle$ \;
    \For{$\delta \in X$}{
        $S^\delta \gets$ \Instantiate{$\mathbf{S}^{\estrobustness}, \delta$} \;
        $\gamma \gets$ \CPSFalsification{$\mathbf{S}^\delta, \mathbf{C}, \phi$} \;
        $v \gets f(\delta, \gamma)$ \;
        $V \gets V \frown \langle v \rangle$ \;
    }
    $\delta_{best} \gets$ \UpdateBest{X, V} \;
    $X \gets$ \NextCandidates{$f, X, V, \estrobustness$} \;
}

\caption{Robustness Falsification Algorithm}
\label{alg:falsification}
\end{algorithm}

Algorithm \ref{alg:falsification} presents the overall process of our proposed falsification framework. It employs a two-layer structure. In the upper layer, the algorithm starts by randomly sample a set of deviations ($X$) from the estimated robustness range $\estrobustness$ (line 2). Then, in the while loop, for each candidate deviation $\delta \in X$, it instantiates a deviated system $S^\delta$ (line 6) and invokes the lower-layer optimization (line 7) to compute the system evaluation value $\gamma$. The value $\gamma$ is then be used to compute the objective function value $v$ (line 8), which are either Eq. \ref{eq:any_deviation} for any violation or Eq. \ref{eq:min_deviation} for minimum violation. Finally, after evaluating all the candidate deviations in this iteration, it updates the best solution so far (line 11) and generates the next candidate deviations based on some heuristics and starts the next iteration (line 12).

Line 7 represents the lower-layer task. It corresponds to our system evaluation function $\Gamma$ as defined in Eq. \ref{eq:system_evaluation}. Therefore, it can be solved by using any CPS falsification method. However, different from traditional falsification tasks which stop on a violation, we let the falsifier to continue finding the minimum STL satisfaction value. This can direct the upper-layer optimization to a region where a violation is more likely to occur.

% In the second layer, given a particular deviation $\delta$, the algorithm instantiates a deviated system $S^\delta$ (line 6) and then solves a CPS falsification problem (line 7). Different from traditional falsification tasks which stop on a violation, we let the falsifier to continue finding the minimum STL satisfaction value. The value $\gamma$ returned by the falsifier will then be used to evaluate the objective function (line 8) which are either formula (\ref{eq:any_deviation}) for any violation or formula (\ref{eq:min_deviation}) for minimum violation.

Our proposed two-layer solving structure provides us the following benefits:
\begin{itemize}
    \item The upper-layer process follows how most evolutionary algorithms would do for optimization. Thus, we are able to integrate many off-the-shelf optimizers into this process. Specifically, we have implemented Uniform Random Sampling, CMA-ES \cite{cmaes}, NSGA-II \cite{nsga2}, Extended Ant Colony \cite{gaco}, and a custom incremental heuristic (which incrementally decreases the boundary for finding minimal deviations).
    \item The lower-layer can be integrated with any CPS falsifier and simulation platform. We have integrated a Python version of CMA-ES and Breach \cite{breach} for falsification, and OpenAI-Gym \cite{OpenAI-Gym}, PyBullet \cite{pybullet}, and Matlab Simulink\footnote{https://www.mathworks.com/products/simulink.html} as the simulation platforms.
    \item Finally, as Avizienis defines robustness as a secondary property \cite{dependability}, this structure separates the concern between robustness and a lower-layer system property such as safety. We could define another system evaluation function $\Gamma$ to evaluate a system. It even gives us the flexibility to replace the classic STL satisfaction function $\rho$ with other computation methods such as cumulative STL \cite{9029429} or mean STL \cite{8814487}. 
\end{itemize}

However, the comparison between different optimization heuristics for the upper layer and different CPS falsifiers for the lower layer is not our main focus. In this paper, we mainly evaluate against Uniform Random Sampling and CMA-ES for the upper layer because: (1) random sampling represents searching without heuristics which is a commonly used baseline method, and (2) from our preliminary experiments, CMA-ES outperforms other evolutionary methods and our incremental heuristic. Therefore, given our limited computation resources, we only run these two methods on our benchmark. Finally, we use CMA-ES as the CPS falsification algorithm as it is a widely used approach \cite{Corso2021-cps-survey,breach} and works most fluently for both Python and Matlab environments.

\section{A CPS Robustness Benchmark}\label{sec:benchmark}
Even though there exist many case study systems from the CPS and RL applications domain, we found that a lot of them do not support configurable parameters to change the system dynamics. Therefore, we present such a benchmark to support our robustness analysis. The benchmark contains eight systems adopted from OpenAI-Gym, PyBullet, and Matlab Simulink. We extend the interfaces of these systems so that users are able to change their behaviors by changing the system parameters. The parameters for systems in our benchmark represent different kinds of deviations including systems' internal deviations, environmental deviations, actuation errors, and observation errors. We also provide classic and RL controllers for these systems so the user is able to compare the robustness between classic and RL controllers. We evaluate our simulation-based analysis framework over this benchmark. However, we believe this benchmark can facilitate a large amount of CPS research not limited to robustness.

\subsubsection{Cart-Pole}
The Cart-Pole problem is described in Section \ref{sec:motivation}. In our experiments, we synthesize a PID and a DQN \cite{Mnih2013-DQN} controller for it; and we define four deviation dimensions, the \emph{Mass of the cart}, the \emph{Mass of the pole}, the \emph{Length of the pole}, and the \emph{Force} when pushing the cart.

\subsubsection{Lunar-Lander}
The Lunar-Lander system\footnote{https://www.gymlibrary.dev/environments/box2d/lunar\_lander/} where the goal is to control an aircraft to safely land on the surface of a planet (within the flagged area). It can fire the main engine (on the bottom) and the left/right engines to control the pose of the aircraft. The safety property defines \emph{1) the rotation of the aircraft should be within a value $\theta$ (e.g., not parallel to the ground or upside down), and 2) it should be close to the landing target as the height decreasing.} In our experiments, we develop a LQR and a PPO \cite{Schulman2017-PPO} controller for it; and we define three deviation dimensions, the \emph{Wind} that can change the x-y position of the aircraft, the \emph{Turbulence} (rotational wind) that can change the rotation of it, and the \emph{Gravity}.

\subsubsection{Car-Circle}
The Car-Circle system where the task is to control a car to move along the circumference of the blue circle \cite{Gronauer2022BulletSafetyGym}. There are ``walls'' on the two sides and the safety property defines \emph{the car should not move across the walls}. In our experiments, we leverage a PPO variation for it from \cite{Liu2023robustness} (which is more robust than standard PPO in the context of robust RL); and we define three types of deviations, the \emph{Force} that moves the car, the \emph{Speed Multiplier} and the \emph{Steering Multiplier} that affect the sensitivity of the forward velocity and the angular velocity response to the force, respectively.

\subsubsection{Car-Run}
The Car-Run system where the task is to control a car move along the track without hitting the walls on the two sides \cite{Gronauer2022BulletSafetyGym}. That is, the safety property defines \emph{the car should not move across the walls}. In our experiments, similar to the Car-Circle system, we also leverage a PPO variation from \cite{Liu2023robustness} and consider the \emph{Speed Multiplier} and the \emph{Steering Multiplier} deviation types.

\subsubsection{Adaptive Cruise Control}
A vehicle equipped with adaptive cruise control (ACC)\footnote{https://www.mathworks.com/help/mpc/ug/adaptive-cruise-control-using-model-predictive-controller.html} has a sensor that measures the distance to the preceding vehicle in the same line. The control goal is to: 1) control the speed of the vehicle to reach the driver-set velocity, and 2) maintain a safe distance to the leading vehicle. Therefore, we have the safety property that \emph{the relative distance between the ego vehicle and the leading vehicle should always be greater than a safe distance}. In our experiments, we adopt an MPC controller from Matlab and a SAC \cite{Haarnoja2018-SAC} controller from Jiayang Song et. al \cite{AI-CPS}. We define three types of deviations, the \emph{Mass} of the vehicle, and the \emph{min} and \emph{max acceleration of the leading vehicle}, changing which can mimic a more progressive or conservative leading vehicle that changes its speed more abruptly or slowly.

\subsubsection{Abstract Fuel Control}
Abstract fuel control (AFC) is a complex air-fuel control system released by Toyota \cite{AFC}. The system takes \emph{PedalAngle} and \emph{EngineSpeed} as input signals and the controller needs to adjust the intake gas rate to the cylinder to maintain optimal air-to-fuel ratio. The safety property is defined such that \emph{the deviation of the air-to-fuel ratio from a reference value should be below a threshold}. We adopt the PID controller from Toyota and a DDPG \cite{Lillicrap2019-DDPG} controller from \cite{AI-CPS}. We adopt the two types of deviations described in \cite{AFC}, \emph{MAF (inlet air mass flow rate) sensor error tolerance} and \emph{Air/Fuel sensor error tolerance}. These are the two error factors in the sensor measurement of MAF and A/F ratio that are used as inputs for the controller. Changing these parameters result in inaccurate state estimation leading to unsafe control actions.

\subsubsection{Lane Keep Assist}
A vehicle equipped with a lane-keeping assist (LKA)\footnote{https://www.mathworks.com/help/mpc/ug/lane-keeping-assist-system-using-model-predictive-control.html} system has a sensor (e.g., a camera) that measures the lateral deviation and relative yaw angle between the center-line of a lane and the vehicle. The sensor also measures the lane curvature and the curvature derivative. The LKA system keeps the car traveling along the center line of the lane by adjusting the steering angle of the ego car. Therefore, the safety property is defined such that \emph{the lateral deviation of the ego vehicle from the center-line of the lane should always be below a threshold}. We adopt an MPC controller from Matlab and a DDPG controller from \cite{AI-CPS}. The Simulink model generates a lane with two turns, and thus we define two types of deviations that change the \emph{Curvature} and the \emph{Position} of the two turns. 

\subsubsection{Water Tank}
A water tank (WTK) system is a container with a controller controlling the inflow and outflow of water, widely used in industry domains like the chemical industry.\footnote{https://www.mathworks.com/help/slcontrol/gs/watertank-simulink-model.html} The safety property is defined such that \emph{the error between the actual water level and the desired water level should always be below a threshold}. We adopt a PID controller from Matlab and a TD3 \cite{Fujimoto2018-TD3} controller from \cite{AI-CPS}. We define two types of deviations, \emph{the water flow rate into the tank} and \emph{the water flow rate out of the tank}, which affect how fast the water volume would change.

\section{Evaluation}\label{sec:evaluation}

\subsection{Research Questions}
We implement our proposed framework in a Python package\footnote{https://figshare.com/s/07274c8faff6de14d0ea} and evaluate our framework through experiments. Specifically, our evaluation focuses on the minimum robustness falsification problem because we believe finding a small deviation closer to the normative environment but leading to a safety violation is more crucial to developers. The evaluation aims to answer the following research questions:
\begin{itemize}
    \item \textbf{RQ1}: Can our proposed robustness falsification framework find small deviations leading to a robustness violation?
    \item \textbf{RQ2}: In terms of minimum robustness falsification, does CMA-ES outperforms Uniform Random search?
\end{itemize}

% Our evaluation focuses on the minimum robustness falsification problem because we believe finding a small deviation that is closer to the normative environment but can lead to a safety violation is more crucial to developers.

% For RQ2, we set Uniform Random as the baseline method for finding robustness violations. Uniform random is a common baseline used in CPS falsification tools \cite{Corso2021-cps-survey}. Since our robustness falsification problem is similar to traditional CPS falsification, so we also use it as our baseline method.
% In addition, we choose CMA-ES because it is an evolutionary algorithm suitable for non-linear non-convex black-box optimization, which is also used in CPS falsification tools like Breach \cite{breach}. Other optimization techniques can be easily integrated into our analysis framework, but a comprehensive comparison across different optimization algorithms is out of the scope of this work.

In addition to these two research questions, we provide observations on the following aspects, even though they are not the main focus of this research:
\begin{itemize}
    \item \textbf{RQ3}: What are the differences between classic controllers and RL controllers in terms of specification-based robustness?
    \item \textbf{RQ4}: For falsification of robustness by optimization, what are the characteristics of the search space? How do they affect the efficacy of the optimization algorithm?
\end{itemize}

The primary research focus of this work is not to compare the robustness of classic and RL controllers and neither to investigate new optimization algorithms for falsification, but these observations might be useful to understand the robustness of RL controllers with respect to changes in system dynamics, different from minor perturbations against a particular model input (e.g., adversarial robustness of ML \cite{Goodfellow2018-adversarial}). Furthermore, understanding the characteristics of the falsification search space can help researchers develop new search heuristics to better find robustness violations or develop new techniques to improve system robustness.

\begin{table*}[!t]
\scriptsize
\centering
\caption{Results from solving the minimum robustness falsification problem. In each run, every problem has a budget of one hundred samples.
% We use the number of samples instead of timeout because different environments take significantly different time to simulate, from seconds to minutes.
A higher \emph{Violation/Total} ratio is better; a lower \emph{Distance} value is better.}
\label{tab:eval-results}
\begin{tabular}{lrrrr}
\toprule
 & \multicolumn{2}{c}{\textbf{Random}} & \multicolumn{2}{c}{\textbf{CMA-ES}} \\ \cmidrule(lr){2-3} \cmidrule(lr){4-5}
 & Violations/Total & Distance (Mean$\pm$Std, Range) & Violations/Total & Distance (Mean$\pm$Std, Range) \\ \midrule
 
\multirow{3}{*}{Cart-Pole-PID} & 11/100 & 0.54$\pm$0.08 (0.38-0.65) & 3/100 & 0.45$\pm$0.03 (0.42-0.48) \\
 & 12/100 & 0.53$\pm$0.07 (0.41-0.64) & 4/100 & 0.56$\pm$0.08 (0.44-0.65) \\
 & 4/100 & 0.58$\pm$0.01 (0.57-0.59) & 10/100 & 0.54$\pm$0.09 (0.43-0.66) \\ \midrule
 
\multirow{3}{*}{Cart-Pole-DQN} & 22/100 & 0.50$\pm$0.07 (0.34-0.65) & 25/100 & 0.47$\pm$0.08 (0.30-0.63) \\
 & 21/100 & 0.46$\pm$0.10 (0.26-0.65) & 37/100 & 0.45$\pm$0.11 (0.28-0.70) \\
 & 23/100 & 0.48$\pm$0.10 (0.28-0.66) & 7/100 & 0.38$\pm$0.04 (0.31-0.44) \\ \midrule
 
\multirow{3}{*}{Lunar-Lander-LQR} & 24/100 & 0.48$\pm$0.10 (0.31-0.64) & 5/100 & 0.24$\pm$0.16 (0.11-0.44) \\
 & 24/100 & 0.47$\pm$0.14 (0.03-0.67) & 13/100 & 0.26$\pm$0.13 (0.04-0.44) \\
 & 25/100 & 0.44$\pm$0.10 (0.24-0.63) & 17/100 & 0.34$\pm$0.14 (0.09-0.62) \\ \midrule
 
\multirow{3}{*}{Lunar-Lander-PPO} & 24/100 & 0.40$\pm$0.15 (0.18-0.63) & 19/100 & 0.26$\pm$0.11 (0.08-0.51) \\
 & 38/100 & 0.42$\pm$0.12 (0.13-0.65) & 25/100 & 0.18$\pm$0.13 (0.03-0.47) \\
 & 37/100 & 0.42$\pm$0.13 (0.05-0.67) & 30/100 & 0.23$\pm$0.13 (0.03-0.52) \\ \midrule
 
\multirow{3}{*}{Car-Circle-PPO} & 29/100 & 0.45$\pm$0.10 (0.23-0.64) & 4/100 & 0.28$\pm$0.10 (0.16-0.40) \\
 & 28/100 & 0.43$\pm$0.11 (0.26-0.69) & 16/100 & 0.21$\pm$0.10 (0.10-0.45) \\
 & 41/100 & 0.40$\pm$0.12 (0.13-0.65) & 2/100 & 0.18$\pm$0.02 (0.16-0.20) \\ \midrule
 
\multirow{3}{*}{Car-Run-PPO} & 33/100 & 0.43$\pm$0.12 (0.22-0.66) & 46/100 & 0.36$\pm$0.13 (0.14-0.68) \\
 & 28/100 & 0.39$\pm$0.13 (0.19-0.65) & 14/100 & 0.31$\pm$0.10 (0.21-0.55) \\
 & 27/100 & 0.42$\pm$0.15 (0.15-0.65) & 16/100 & 0.28$\pm$0.10 (0.15-0.46) \\ \midrule
 
\multirow{3}{*}{ACC-SAC} & 23/100 & 0.45$\pm$0.12 (0.15-0.68) & 8/100 & 0.22$\pm$0.06 (0.13-0.32) \\
 & 17/100 & 0.42$\pm$0.09 (0.25-0.59) & 6/100 & 0.21$\pm$0.10 (0.11-0.42) \\
 & 15/100 & 0.47$\pm$0.12 (0.21-0.65) & 29/100 & 0.37$\pm$0.12 (0.17-0.59) \\ \midrule
 
\multirow{3}{*}{ACC-MPC} & 0/100 & - & 1/100 & 0.70$\pm$0.00 (0.70-0.70) \\
 & 1/100 & 0.66$\pm$0.00 (0.66-0.66) & 3/100 & 0.67$\pm$0.03 (0.64-0.71) \\
 & 0/100 & - & 1/100 & 0.64$\pm$0.00 (0.64-0.64) \\ \midrule
 
\multirow{3}{*}{AFC-DDPG} & 0/100 & - & 0/100 & - \\
 & 0/100 & - & 0/100 & - \\
 & 0/100 & - & 0/100 & - \\ \midrule
 
\multirow{3}{*}{AFC-PID} & 0/100 & - & 0/100 & - \\
 & 0/100 & - & 0/100 & - \\
 & 0/100 & - & 0/100 & - \\ \midrule
 
\multirow{3}{*}{WTK-TD3} & 24/100 & 0.52$\pm$0.07 (0.39-0.69) & 9/100 & 0.48$\pm$0.03 (0.43-0.52) \\
 & 29/100 & 0.49$\pm$0.08 (0.34-0.61) & 15/100 & 0.44$\pm$0.06 (0.34-0.55) \\
 & 37/100 & 0.50$\pm$0.07 (0.37-0.68) & 21/100 & 0.37$\pm$0.07 (0.30-0.57) \\ \midrule
 
\multirow{3}{*}{WTK-PID} & 37/100 & 0.41$\pm$0.12 (0.17-0.62) & 17/100 & 0.26$\pm$0.15 (0.08-0.59) \\
 & 33/100 & 0.40$\pm$0.14 (0.12-0.65) & 24/100 & 0.14$\pm$0.07 (0.06-0.31) \\
 & 38/100 & 0.41$\pm$0.12 (0.21-0.69) & 20/100 & 0.21$\pm$0.10 (0.08-0.43) \\ \bottomrule
\end{tabular}
\end{table*}

\subsection{Experiment Results}
To answer these research questions, we run the minimum robustness falsification problem (Problem \ref{prob:min-robustness-falsify}) on our robustness benchmark\footnote{LKA is not included because it requires too many resources to run that we cannot complete in an acceptable time.}. For analyzing and illustration purpose, we show the experiment results for systems with two deviation dimensions. All experiments run on a Linux machine with a 3.6GHz CPU and 24GB memory. Each system+controller combination is ran three times using CMA-ES and Uniform Random Search, respectively.

Table \ref{tab:eval-results} summarizes the results for solving the minimum robustness falsification problems. For each run, we set a 100-sample budget for both Random and CMA-ES; the \emph{Violations/Total} column shows the number of violations found out of the one hundred samples. The \emph{Distance} column shows the average normalized $l\text{-2}$ norm to the zero-deviation point (i.e., $\lVert \delta - \delta_0 \rVert_2$) of the found violations, with its standard deviation and range.

From the table, Random and CMA-ES can both find violations for all problems except AFC. Random can often find more violations than CMA. However, in ACC with a MPC controller, Random only finds one violation from three runs but CMA-ES finds violations in all three runs. On the other hand, while CMA-ES finds fewer violations, the average distance of the violations are significantly smaller than Random, with one exception in Cart-Pole. In addition, since CMA-ES is thought to be more effective with high-dimensional variables, we conduct the same experiments for systems which support more than two deviation dimensions. We obtain similar results which are described in the Appendix.

\subsection{Analysis of Results}
\begin{figure*}[!t]
    \centering
    \subfloat[Cartpole-PID]{
        \includegraphics[width=0.22\linewidth]{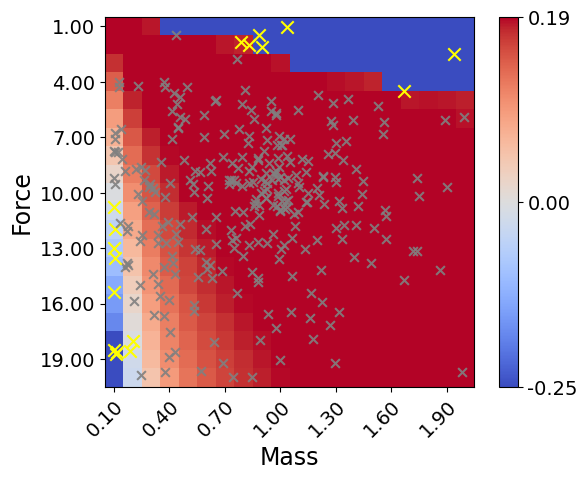}
    }
    % \hfill
    % \subfloat[Cartpole-PID with Random]{
    %     \includegraphics[width=0.2\linewidth]{figures/cartpole-pid/fig-violations-random-all.png}
    % }
    \hfill
    \subfloat[Cartpole-DQN]{
        \includegraphics[width=0.22\linewidth]{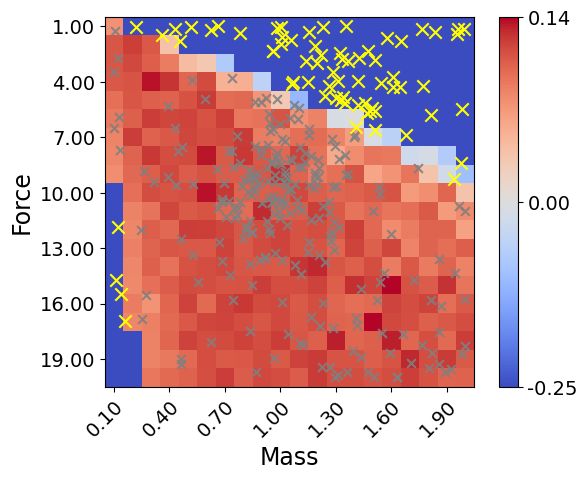}
    }
    % \hfill
    % \subfloat[Cartpole-DQN with Random]{
    %     \includegraphics[width=0.2\linewidth]{figures/cartpole-dqn/fig-violations-random-all.png}
    % }
    \hfill
    \subfloat[Lunar-Lander-LQR]{
        \includegraphics[width=0.22\linewidth]{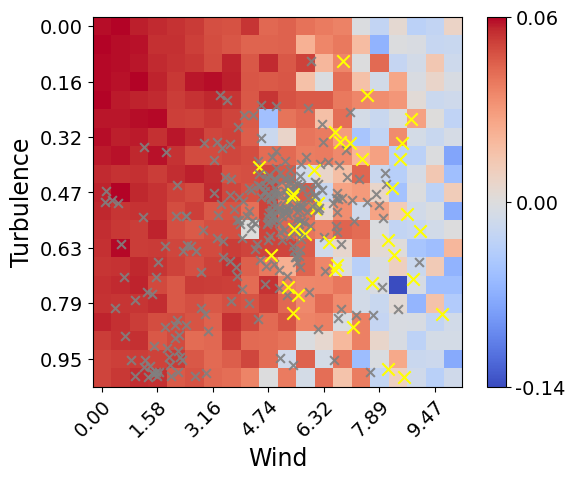}
    }
    % \hfill
    % \subfloat[Lunar-Lander-LQR with Random]{
    %     \includegraphics[width=0.2\linewidth]{figures/lunar-lqr/fig-violations-random-all.png}
    % }
    \hfill
    \subfloat[Lunar-Lander-PPO]{
        \includegraphics[width=0.22\linewidth]{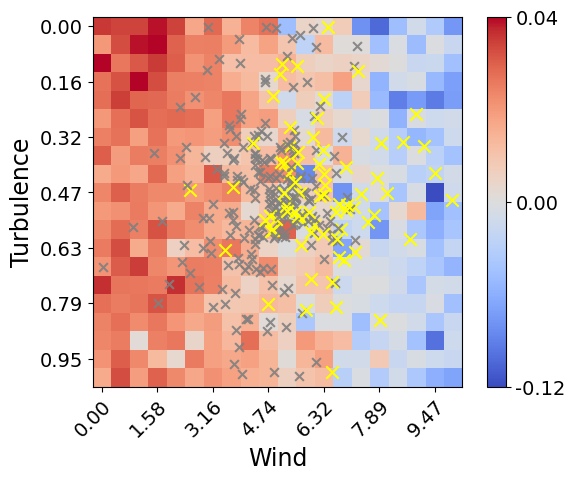}
    }
    % \hfill
    % \subfloat[Lunar-Lander-PPO with Random]{
    %     \includegraphics[width=0.2\linewidth]{figures/lunar-ppo/fig-violations-random-all.png}
    % }
    \hfill
    \subfloat[Car-Run-PPO]{
        \includegraphics[width=0.22\linewidth]{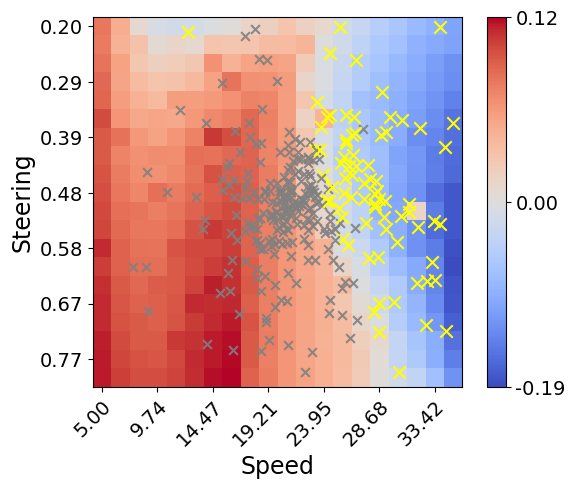}
    }
    % \hfill
    % \subfloat[Car-Run-PPO with Random]{
    %     \includegraphics[width=0.2\linewidth]{figures/car-run/fig-violations-random-all.png}
    % }
    \hfill
    \subfloat[Car-Circle-PPO]{
        \includegraphics[width=0.22\linewidth]{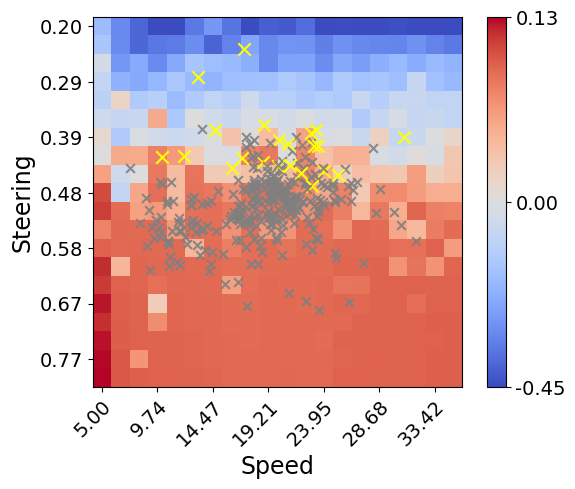}
    }
    % \hfill
    % \subfloat[Car-Cicle-PPO with Random]{
    %     \includegraphics[width=0.2\linewidth]{figures/car-circle/fig-violations-random-all.png}
    % }
    \hfill
    \subfloat[ACC-MPC]{
        \includegraphics[width=0.22\linewidth]{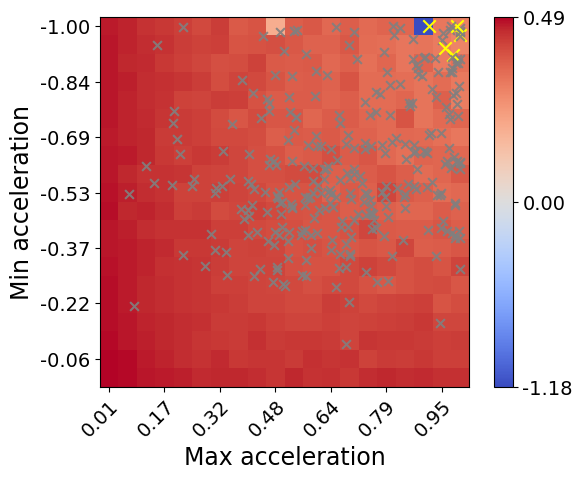}
    }
    % \hfill
    % \subfloat[ACC-MPC with Random]{
    %     \includegraphics[width=0.2\linewidth]{figures/ACC-T/fig-violations-random-all.png}
    % }
    \hfill
    \subfloat[ACC-SAC]{
        \includegraphics[width=0.22\linewidth]{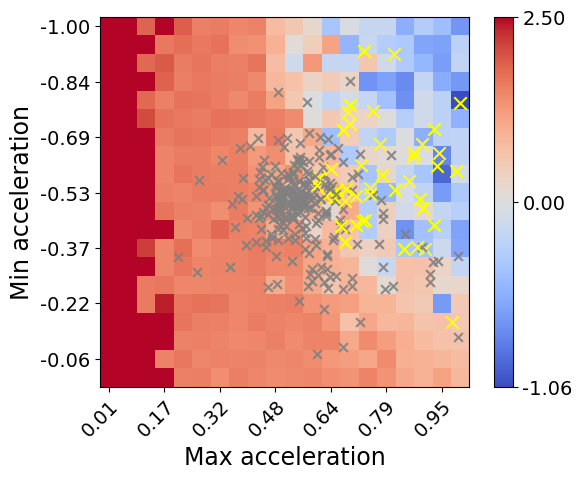}
    }
    % \hfill
    % \subfloat[ACC-SAC with Random]{
    %     \includegraphics[width=0.2\linewidth]{figures/ACC-RL/fig-violations-random-all.png}
    % }
    \hfill
    \subfloat[AFC-PID]{
        \includegraphics[width=0.22\linewidth]{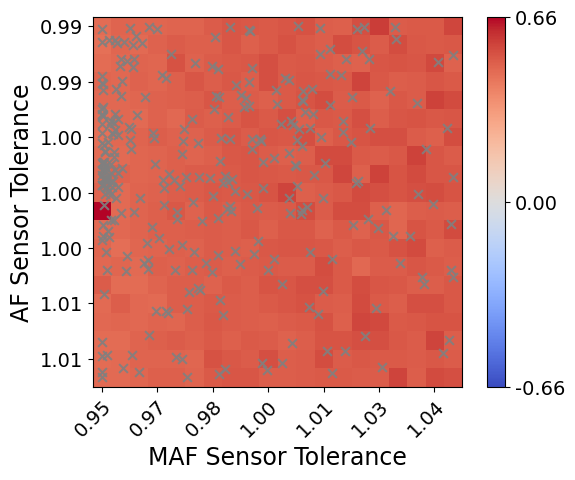}
    }
    % \hfill
    % \subfloat[AFC-PID with Random]{
    %     \includegraphics[width=0.2\linewidth]{figures/AFC-T/fig-violations-random-all.png}
    % }
    \hfill
    \subfloat[AFC-DDPG]{
        \includegraphics[width=0.22\linewidth]{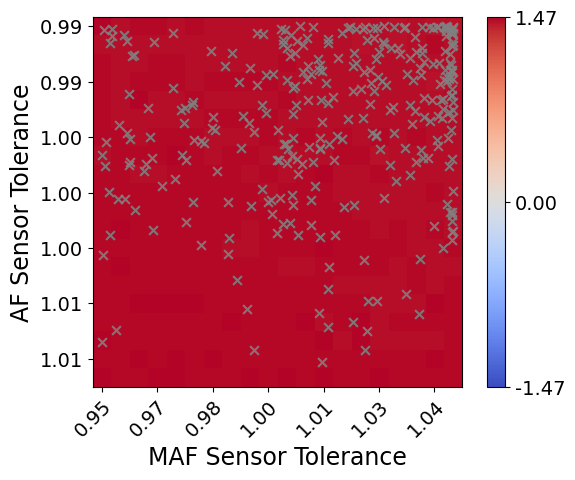}
    }
    % \hfill
    % \subfloat[AFC-DDPG with Random]{
    %     \includegraphics[width=0.2\linewidth]{figures/AFC-RL/fig-violations-random-all.png}
    % }
    \hfill
    \subfloat[WTK-PID]{
        \includegraphics[width=0.22\linewidth]{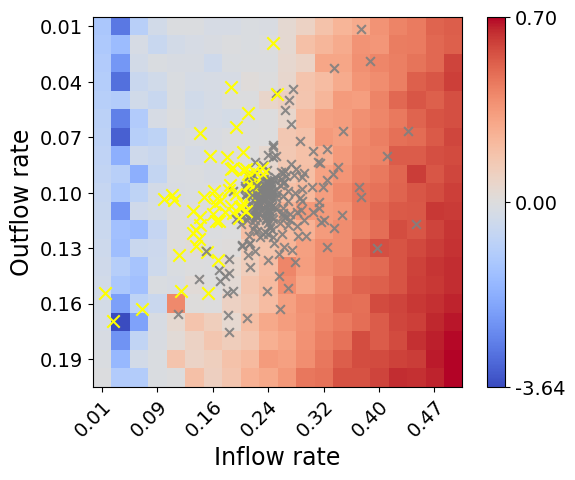}
    }
    % \hfill
    % \subfloat[WTK-PID with Random]{
    %     \includegraphics[width=0.2\linewidth]{figures/WTK-T/fig-violations-random-all.png}
    % }
    \hfill
    \subfloat[WTK-TD3]{
        \includegraphics[width=0.22\linewidth]{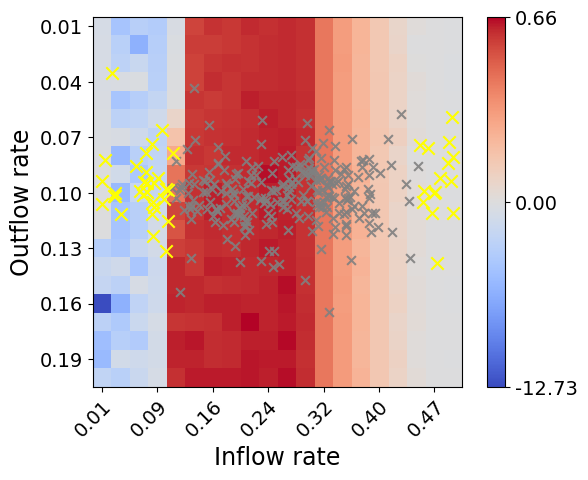}
    }
    % \hfill
    % \subfloat[WTK-TD3 with Random]{
    %     \includegraphics[width=0.2\linewidth]{figures/WTK-RL/fig-violations-random-all.png}
    % }
    \caption{Search space and samples by CMA-ES. A red point on the heat map signifies a positive STL satisfaction value, and a blue point signifies a negative value; the intensity of the color corresponds to the magnitude of this value. A cross ($\times$) on the heat map indicates a sampled deviation, and a yellow cross indicates a violating deviation. }
    \label{fig:heatmaps}
\end{figure*}

To better understand the experiment results (e.g., why CMA-ES finds fewer violations, why it does not outperform Random in Cart-Pole), we plot the problem search spaces in heat maps and plot the samples and the violations as dots. Each heat map is generated by slicing the search space (i.e., the estimated range of system parameters) into a grid and using a CPS falsifier to find the minimum STL satisfaction value for each grid cell. Specifically, we use a $20\times20$ grid resulting in 400 CPS falsification calls, far more than our budget for solving the robustness falsification problem. Figure \ref{fig:heatmaps} shows the heat maps for all the problems with CMA-ES.

% \footnote{Diagrams for Random are omitted because the background heat maps are the same, and the samples are evenly scattered in the search spaces as we apply uniform sampling.}
\paragraph{\textbf{Answer to RQ1}} From Table \ref{tab:eval-results} and Figure \ref{fig:heatmaps}, our framework with both Random search and CMA-ES can find violations to robustness. They both failed in AFC because it is always robust in the parameter range given by Toyota. Then, by solving the minimum robustness falsification problem, violations found by CMA-ES are small enough which successfully converge to the safe and unsafe boundary.

\paragraph{\textbf{Answer to RQ2}} From the table, we can observe a significant difference in the distance metric of the violations found by Random and CMA-ES. Also, from the heat maps, violations found by CMA-ES would converge to the safe-unsafe boundary, while violations by Random are scattered in the space. Nevertheless, due to this convergence, CMA-ES is looking for samples around the boundary causing it to find fewer violations. Moreover, when a system is mostly robust with only a small unsafe region (e.g., ACC-MPC), Random would fail to find a violation. In sum, we conclude that CMA-ES can more efficiently find minimum robustness violations.

\paragraph{\textbf{Observations for RQ3}} From the diagrams, we observe that classic controllers are better than RL controllers in terms of area of safe regions and max STL satisfaction values (e.g., in Cart-Pole, Lunar-Lander, and ACC). Also, although the shapes of the search spaces vary significantly across different problems,
% (which are highly affected by system dynamics, deviations, and controllers)
we can often observe a ``smooth'' descending trend from a high STL satisfaction region to a low satisfaction region in classic controllers, whereas the change is more abrupt in RL controllers. For example, in Cart-Pole-DQN, the satisfaction value exhibits local fluctuations around a deviation point; in ACC-SAC, the value abruptly drops at \emph{Max acceleration $\approx 0.17$}; and in WTK-TD3, the value suddenly drops from positive to negative near \emph{Inflow rate $\approx 0.1$}.
% Such ``smoothness'' of change might be captured by continuity properties such as Lipschitz continuity.

\paragraph{\textbf{Observations for RO4}} The shape of the search space highly depends on the system dynamics, controllers, and deviation types. There does not exist a uniform pattern among these problems, which meet our intuition that a black-box second-order method is necessary. We find that CMA-ES as a derivative-free optimization method can successfully converge to the robust (positive) and unrobust (negative) boundary of the space where a minimum robustness violation should be located.
% Since it is looking for samples around this boundary, it is reasonable that it finds fewer violations.
On the other hand, it does not perform well particularly in the Cart-Pole-PID problem because there is an abrupt decrease at the boundary (top-right) and the robust region is flat (i.e., the system is always at the highest STL satisfaction level around that region), making it hard to converge. Therefore, a ``smooth'' change of the search space would also benefit the performance of such evolutionary methods.

Although further investigation needs to be done to validate these observations for RQ3\&4, it also points out an interesting research direction: If it's too hard to build a controller that is consistently robust in all deviations, how we can build a controller that can gradually and smoothly degrade in performance given increasing deviations? Such ``smoothness'' of change might be captured by continuity properties such as Lipschitz continuity. In other words, the degradation with respect to certain specification should be more predictable without abrupt changes, especially for RL controllers.

% \subsection{Discussion}

% \begin{tcolorbox}[title=\textbf{Answer to Research Questions:}]
% \textbf{RQ1:} Our tool can be applied to various systems (using different simulation platforms) and controllers. Both Random and CMA-ES can find violations to robustness. \ag{Is it surprising, and worth highlighting that the first layer Uniform Random sampling can find violations?}\cj{If we set a large deviation space, then it won't be too surprise that random should work.}
% \tcblower
% \textbf{RQ2:} Uniform random is able to find more violations than CMA. However, in terms of minimum robustness falsification, violations found by CMA-ES have smaller average distances than Random. On the other hand, if only very few violations exist in the search space, Random may not find any violation but CMA-ES can.
% \end{tcolorbox}

% \begin{tcolorbox}[title=\textbf{Insights:}]
% \textbf{Insight1:} Classic controllers behave more smoothly than RL controllers w.r.t. system parameter changes in that they have fewer abrupt changes in the STL satisfaction value.
% \tcblower
% \textbf{Insight2:} Although derivative-free optimization algorithms like CMA-ES do not rely on derivatives, a search space with a ``smoother'' change in its function value (e.g., satisfying Lipschitz continuity) may still help it converge, which facilitates such robustness analysis.
% \end{tcolorbox}

\subsection{Threats to Validity}
Our evaluation focuses on solving the minimum robustness falsification problem. The comparison between Random and CMA-ES and the observations based on the heat maps refer to the upper-layer optimization described in Algorithm \ref{alg:falsification}. However, the validity of the results also rely on the quality of the lower-layer CPS falsification, where its computation may not stably find the minimum STL satisfaction value. We mitigate the threat by using the same falsification algorithm with the same configuration for all problems with the same searching budget.

% \todo{Parameter identification can be it's own subproblem and some parameters might be easier to investigate/falsify due to smooth optimisation curves. parameter+specification+policy define the loss curve.}\cj{Not sure if this should be in threats to validity}

\section{Related Work}\label{sec:related_work}
Falsification of CPS \cite{Corso2021-cps-survey} is a well-studied problem in the literature, the most mentioned ones are Breach \cite{breach} and PSY-TaLiRo \cite{Thibeault2021-psytaliro}. These tools mutate the initial states or inputs to a system to find counterexamples that violate certain STL property. Compared to traditional CPS falsification problem, our robustness falsification problem focuses on a high-layer of abstraction that we aim to find system parameters that would lead to a specification violation.

VerifAI \cite{VerifAI,Sanjit-compositional} applies a similar idea that they consider \emph{abstract features} for a ML model that can lead to a property violation of a CPS. Different from us, they assume a CPS with a ML perception model (such as object detection) connecting to a classic controller, and the abstract features are environmental parameters that would affect the performance of the ML model (e.g., brightness). In other words, they focus on deviations that affect the ML model whereas our deviation notion is more general that includes any external or internal deviation or sensor error which changes the system dynamics.

Although the problem is different, we share similarities with these tools in the falsification methods, i.e., we all apply sampling-based optimization algorithms to find violations. One key difference is that we adopt a two-layer search structure which explicitly distinguishes between system parameters and system initial states and inputs. This structure separates the concern between robustness and a lower-layer system property such as safety and provides us flexibility to adopt other lower-layer system evaluation techniques.
% As Avizienis defines robustness as a secondary property \cite{dependability}, this structure separates the concern between robustness and a lower-layer property such as safety. It also gives us the flexibility to adopt other techniques to evaluate a system with respect to another lower-layer property.

Robust RL studies the problem to improve the performance of RL agents in the presence of uncertainties (including distribution shifts, dynamic uncertainties, action perturbations, and observation errors) and adversarial attacks \cite{Moos-RobustRL, xu2022trustworthy}. A similar research topic is domain randomization \cite{Peng2018-Sim-to-Real,Sadeghi2017cad2rl,Tobin2017domain} that create various systems with randomized properties leading to changed system dynamics and then train a model that works across these systems. However, our work is different from these works in that: (1) we focus on robustness evaluation whereas they focus more on training; and (2) we focus on system specifications and specify them in STL properties, while they rely on rewards where maximizing the reward function does not necessarily guarantee certain system specification. 

% \cj{2. Work mentioned by Parv on mutating system parameters for testing}

\section{Conclusion}\label{sec:conclusion}
In this paper, we have introduced \emph{specification-based robustness}, where a controller is said to be robust if it is capable of ensuring a desired system requirement (specified in STL) even when the system deviates from its nominal dynamics. This definition yields a new type of analysis problem, called  \emph{robustness falsification}, where the goal is to find small changes to the system dynamics that result in a violation of a given STL specification. We have also presented a novel optimization-based approach to solve the robustness falsification problem, and evaluated the effectiveness of the approach over a set of benchmark CPS problems. 

In our current analysis, we employ the strict robustness definition, i.e., any trajectory of a (deviated) system should satisfy the desired STL specification. However, this is a rather restrictive guarantee that ignores the probability distributions of a stochastic system. Some trajectories might be extremely unlikely to occur, and it might be less critical and cost-effective to make the controller be robust against them. In such a case, computing the expected STL satisfaction value of a stochastic system might be more practical. Note that our analysis framework allows changing the system evaluation function (Eq. \ref{eq:system_evaluation}). As part of future work, we plan to explore and integrate other types of evaluation functions $\Gamma$ or even different semantics of STL satisfaction function $\rho$ (e.g., cumulative robustness\cite{9029429}).

To measure the distance between nominal and deviated systems, we use normalized $l\text{-2}$ norm over system parameters. However, this measure cannot capture certain types of changes in system dynamics. For example, in the Cart-Pole, if the mass and the force are modified at the same time, this would result in a possibly large $l\text{-2}$ distance, but the overall system dynamics may remain similar to the nominal one. We plan to explore other types of distance measures, such as Wasserstein Distance \cite{Lecarpentier2019-wasserstein,Abdullah2019-wasserstein,Yang2017-wasserstein}, which captures the distance in distributions and can potentially better measure the robustness of a controller with respect to changes in system dynamics.

\bibliographystyle{IEEEtran}
\bibliography{IEEEabrv,ref}

\clearpage

\section*{Appendix}
\subsection{Results from multi-deviation-dimension experiments}
Table \ref{tab:eval-results-2} shows the results for solving the minimum robustness falsification problems for systems with more than two deviation dimensions. We observe a similar phenomenon compared to the two-dimensional case, i.e., Uniform Random can often find more violations where as CMA-ES is able to find smaller-distance violations. Also, in the case where there are very few violations, Random may not be able to find a counterexample but CMA-ES can.

\begin{table*}[!t]
\centering
\caption{Results of the minimum robustness falsification problem with more than two dimensions.}
\label{tab:eval-results-2}
\begin{tabular}{lrrrr}
\toprule
 & \multicolumn{2}{c}{\textbf{Random}} & \multicolumn{2}{c}{\textbf{CMA}} \\ \cmidrule(lr){2-3} \cmidrule(lr){4-5}
 & Violations/Total & Distance (Mean$\pm$Std, Range) & Violations/Total & Distance (Mean$\pm$Std, Range) \\ \midrule
 
\multirow{3}{*}{Cart-Pole-4-PID} & 18/100 & 0.63$\pm$0.09 (0.46-0.79) & 0/100 & - \\
 & 12/100 & 0.65$\pm$0.11 (0.46-0.82) & 8/100 & 0.70$\pm$0.06 (0.60-0.76) \\
 & 10/100 & 0.65$\pm$0.06 (0.55-0.76) & 48/100 & 0.54$\pm$0.08 (0.38-0.79) \\ \midrule
 
\multirow{3}{*}{Cart-Pole-4-DQN} & 24/100 & 0.60$\pm$0.11 (0.29-0.75) & 12/100 & 0.46$\pm$0.09 (0.28-0.59) \\
 & 22/100 & 0.62$\pm$0.13 (0.37-0.82) & 19/100 & 0.44$\pm$0.08 (0.31-0.66) \\
 & 28/100 & 0.56$\pm$0.11 (0.28-0.72) & 8/100 & 0.45$\pm$0.14 (0.27-0.68) \\ \midrule
 
\multirow{3}{*}{Lunar-Lander-3-LQR} & 41/100 & 0.70$\pm$0.18 (0.16-0.98) & 41/100 & 0.33$\pm$0.11 (0.18-0.80) \\
 & 45/100 & 0.67$\pm$0.18 (0.24-0.96) & 48/100 & 0.31$\pm$0.10 (0.10-0.55) \\
 & 48/100 & 0.71$\pm$0.17 (0.21-1.00) & 50/100 & 0.41$\pm$0.12 (0.14-0.63) \\ \midrule
 
\multirow{3}{*}{Lunar-Lander-3-PPO} & 73/100 & 0.67$\pm$0.18 (0.28-1.01) & 51/100 & 0.39$\pm$0.11 (0.17-0.60) \\
 & 71/100 & 0.66$\pm$0.18 (0.17-0.99) & 48/100 & 0.35$\pm$0.13 (0.08-0.65) \\
 & 66/100 & 0.63$\pm$0.20 (0.12-1.00) & 26/100 & 0.38$\pm$0.18 (0.03-0.79) \\ \midrule
 
\multirow{3}{*}{Car-Circle-3-PPO} & 37/100 & 0.64$\pm$0.16 (0.28-0.98) & 19/100 & 0.28$\pm$0.09 (0.12-0.48) \\
 & 36/100 & 0.60$\pm$0.20 (0.33-0.99) & 15/100 & 0.31$\pm$0.09 (0.16-0.45) \\
 & 30/100 & 0.62$\pm$0.20 (0.14-1.00) & 10/100 & 0.26$\pm$0.07 (0.17-0.40) \\ \midrule
 
\multirow{3}{*}{ACC-3-MPC} & 0/100 & - & 2/100 & 0.68$\pm$0.03 (0.65-0.71) \\
 & 1/100 & 0.76$\pm$0.00 (0.76-0.76) & 2/100 & 0.65$\pm$0.00 (0.65-0.65) \\
 & 0/100 & - & 1/100 & 0.77$\pm$0.00 (0.77-0.77) \\ \midrule
 
\multirow{3}{*}{ACC-3-SAC} & 24/100 & 0.45$\pm$0.13 (0.12-0.69) & 5/100 & 0.30$\pm$0.10 (0.18-0.44) \\
 & 18/100 & 0.48$\pm$0.10 (0.21-0.66) & 13/100 & 0.19$\pm$0.07 (0.09-0.31) \\
 & 17/100 & 0.56$\pm$0.10 (0.40-0.79) & 25/100 & 0.31$\pm$0.12 (0.13-0.58) \\ \bottomrule
\end{tabular}
\end{table*}

\end{document}